\documentclass[aps,prb,twocolumn,superscriptaddress,showpacs]{revtex4}
\usepackage{amsmath}
\usepackage{amssymb}
\usepackage{graphicx}
\usepackage{subfigure}
\usepackage{epstopdf}
\usepackage[usenames,dvipsnames]{color}
\usepackage{float}
\begin{document}
\title{Entanglement spectrum edge reconstruction and correlation hole of the FQH liquids}
\author{Li-Xia Wei}
\affiliation{Department of Physics, Chongqing University, Chongqing 401331, P. R. China}
\author{Na Jiang}
\affiliation{Zhejiang Institute of Modern Physics, Zhejiang University, Hangzhou 310027, P. R. China}
\author{Qi Li}
\affiliation{Department of Physics, South University of Science and Technology of China, Shenzhen 518055, P. R. China} 
\affiliation{School of Physics and Technology, Wuhan University, Wuhan 430072, P. R. China}
\author{Zi-Xiang Hu}
\email{zxhu@cqu.edu.cn}
\affiliation{Department of Physics, Chongqing University,Chongqing 401331, P. R. China}

\date{\today}
\begin{abstract}

The edge of the electronic fractional quantum Hall (FQH) system obeys the law of the chiral Luttinger liquid theory due to its intrinsic topological properties and the relation of bulk-edge correspondence. However, in a realistic experimental system, such as the usual Hall bar setup, the soften of the background confinement potential can induce the reconstruction of the edge spectrum which breaks the chirality and universality of the FQH edge. The entanglement spectrum (ES) of the FQH ground state has the same counting structure as that in the energy spectrum indicating the topological characters of the quantum state.  In this work, we report that the ES can also have an edge reconstruction while sweeping the area of the sub-system in real space cut. Moreover, we found the critical area of the sub-system matches accurately with the intrinsic building block of the fractional quantum Hall liquids, namely the correlation hole of the FQH liquids. The above results seem like be universal after our studying a series of typical FQH states, such as two Laughlin states at $\nu = 1/3$ and $\nu = 1/5$, and the Moore-Read state for $\nu = 5/2$.

\end{abstract}
\maketitle

\section{Introduction}
Topological states of matter have been a major theme in the recent developments in understanding novel quantum effects. The inherent fractional quantum Hall (FQH)  effects~\cite{Tsui,Laughlin83}  in a high mobility two-dimensional electron  gases (2DEGs)  under a strong magnetic field is a significant field to explore the topological phases. In these cases, the bulk of a FQH liquid is gapped which gives  incompressibility of the topological quantum ground state.~\cite{Girvin, Jainbook} However, the low-lying gapless excitations exist at the boundary of the liquid that provides a unique arena to study electron correlations in one dimension and topological properties in the bulk due to the bulk-edge correspondence in topological phases. It is known that topological states characteristically have protected edge states at the boundary between trivial and non-trivial regions. On the quantum Hall edge, the Fermi liquid theory breaks down and the edge electrons have been argued ~\cite{WenIJPM} to form the chiral Luttinger liquid (CLL).  The CLL theory predicted that the current-voltage dependence in the tunneling between a Fermi liquid and a quantum Hall edge follows a universal power-law $I \sim V^{\alpha}$ where $\alpha = m$ for the $\nu = \frac{1}{m}$ FQH states.~\cite{Wen94,ChangRMP03} Such universality, however, has a long time controversy since it has not been conclusively observed in semiconductor-based 2DEGs in spite of that the graphene-based 2DEGs has possibility to realize this universality~\cite{huPRL,Andrei12}. One possible reason of this discrepancy is edge reconstruction~\cite{Wen94,ChangRMP03}, which induces additional non-chiral edge modes that are not tied to the bulk topology. Edge reconstruction is a consequence of competition between the confinement potential that holds the electrons in the interior of the sample, and Coulomb repulsion that tends to spread out the electron density.  In numerical studies~\cite{Wan02, Wan03, HuPRB16, Jain09, Jain10},  the knob of the edge reconstruction was the distance between the 2DEGs and background potential, or the strength of the confinement potential.

Another aspect to explore the edge excitation of the FQH state is the entanglement spectrum (ES)~\cite{HaldaneES} of the ground state wave function. The ES is the ``energy spectrum" of the bipartite reduced density matrix of the ground state. It has a deep connection to the topological properties embedded in the ground state and its low-lying excitations. This connection is based on the conformal field theory (CFT) description of the FQH model wave functions, such as the Abelian Laughlin states at filling $\nu  = 1 / m$ and the non-Abelian Read-Rezayi states~\cite{RR} with order-$k$ clustering at filling $\nu = \frac{k}{kM+2}$. Mostly, ES reveals the bulk topology via counting the low energy excited states which are regarded as the virtual edge excitation at the boundary of the bipartition. In analogy to the electron energy spectrum of an open boundary system,  the counting numbers of the ES for Laughlin state are ``$1,1,2,3,5 \cdots$ and  ``$1,1,3,5,10 \cdots$ for Moore-Read state, which can be predicted by CFT or CLL theory. However, except the counting numbers, there is few study of the quantitative properties of the ES, such as its edge velocities or edge reconstruction and their relation to the bulk topology. This is because these properties are mostly believed to be non-universal and unrelated to the bulk topology.

The real space bipartite cut~\cite{Regnault12,Dubail12, Simon12} has advantage that its unambiguously determining the boundary between two subsystems which is helpful for quantification of the boundary length and subsystem area. Therefore, the real space bipartition is the best way to fit the ``area law" of the entanglement entropy and extrapolate the topological entanglement entropy.~\cite{Zanardi, Kitaev, Levin} In this paper,  we systematically study the properties of the real space ES of the FQH states in disk geometry. We find that the reconstruction of the ES can be realized by continuously tuning the position of the cut, namely the area of the subsystem with a fixed number of electrons in subsystem. Interestingly, the critical area of the subsystem exactly matches the elementary unit of the fractional quantum Hall liquids. For example, the essential unit of the $1/3$ Laughlin state is the electron bound to a correlation hole corresponding to ``units of flux", or three of the available single particle states which are exclusively occupied by the particle to which they are attached. In general, the elementary unit of the FQH liquid is a ``composite boson" of $p$ particles with q attached quanta which is an analog of unit cell in a solid. This conclusion is supported by our series of verifications, such as the model wave functions for Laughlin state at $\nu = 1/3$ and $\nu = 1/5$ and Moore-Read state~\cite{MR} at $\nu = 5/2$ and their counterparts for Coulomb interaction. 

The rest of this paper is organized as following: In Sec. II, the model and the edge reconstruction are reviewed. The ES spectrum, its reconstruction, edge velocity and subsystem entanglement entropy for the $1/3$ Laughlin state are studied in Sec. III.  The verifications for the realistic Coulomb interaction and other FQH states are considered in Sec. V and Sec. IV gives the conclusions and discussions.
   
\section{Review of the model and edge reconstruction in energy spectrum}

We consider a semi-realistic microscopic model for FQH liquids in GaAl/GaAlAs heterostructure which contains the 2DEGs layer locates at the interface between GaAs and GaAlAs, and a uniform distributed positive background attribute to the dopants at a distance $d$.  The background confinement competes with the electron-electron interaction, which is the driven force of the edge reconstruction. The density of the background charge equals to the filling factor $\sigma = \nu$ and its overall charge cancels the charge of the electrons due to the charge neutrality condition.  Therefore, the background potential is a single body potential in the FQH problem.  In case of without considering the Landau level (LL) mixing and the spin degree of freedom for simplicity (the ground state is supposed to be spin polarized), the Hamiltonian can be written as: 
\begin{equation}
\centering
\label{Hamiltonian}
 H = \frac{1}{2}\sum_{\{m_i\}}V_{1234}c_{m_1}^\dagger c_{m_2}^\dagger c_{m_3}c_{m_4} + \sum_m U_m c_m^\dagger c_m,
\end{equation}
where the $c_m^\dagger$ is the electron creation operator for the lowest Landau level (LLL) single electron state $\phi_m=|m\rangle$ with angular momentum $m$. The matrix element of two-body interaction and background potential are:
\begin{equation}
\centering
 V_{1234} = \langle m_1,m_2|V(r_1-r_2)|m_3,m_4\rangle,
\end{equation}
and 
\begin{equation}
\centering
U_m = e\sigma \int_{r_2 < R} d^2r_2 \langle m|\frac{1}{\sqrt{d^2 + |\vec{r_1}-\vec{r_2}|^2}}|m\rangle.
\end{equation}
The advantage of this model is that, by tuning the parameters $d$, the FQH phases and their reconstructed edge states, emerge naturally as the global ground state of the microscopic Hamiltonian without any explicit assumptions, e.g., on the value of the ground state angular momentum. Thus we can study the stability of the phases and their competitions. Another advantage of the model is that we can analyze the edge excitations of the semi-realistic system and identify them in a one-to-one correspondence with CLL edge theory or CFT. In addition to confirm the bulk topological order, we can use the microscopic calculation to extract energetic quantities, such as edge velocities, which are crucial for quantitative comparisons with experiments. The distance $d$ between the 2DEGs and the uniform background potential is the parameter which tunes the relative strength between electron-electron  repulsion and the attraction from positive background. When $d$ is small, the confinement is strong and electrons tend to stay in the interior of the sample, however, the confinement becomes weaker while increasing $d$ and thus the electrons especially the ones near the edge, can spread out. This is the main mechanism of the edge reconstruction.  
\begin{figure}[H]
\includegraphics[width=8.5cm]{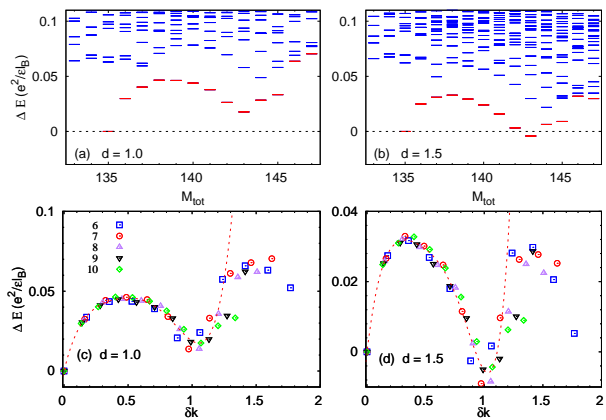}
\caption{\label{fig1}The Coulomb energy spectrum for 10 electrons in 30 orbitals before (a) and after (b) edge reconstruction in disk geometry while tuning the background potentail parameterized by $d$. The total angular momentum for the global ground state has an abrupt change from the Laughlin-like state with $M_{tot} =M_0 = 3N(N-1)/2 = 135$ to $M_{tot} = 143$ at the critical value of $d_c \sim 1.5l_B$. (c) and (b) depict the dispersion relations of the edge mode in (a) and (b) for different system sizes.}
\end{figure}

After projecting to the LLL, the effective interaction of a two-body interaction can be expanded by a set of  Haldane's pseudopotentials $\{V_m\}$.~\cite{HaldanePP} It is  known that the Laughlin state at $\nu = 1/3$: 
\begin{equation}\label{lauglinwf}
\Psi_L^3(\{z_i\})=\prod_{i<j}(z_i-z_j)^3 e^{-\frac{1}{4}\sum_i |z_i|^2}
\end{equation}
is the exact zero energy eigenstate for $V_1$ model Hamiltonian. Another way of obtain the model wave function is the Jack polynomials. It is known~\cite{Bernevig1, Bernevig2} that the FQH wave functions can be calculated recursively by Jacks with a negative parameter $\alpha$ and a root configuration.  The root configuration satisfies  $(k, r)$ admissibility which means there can be at most $k$ particles in $r$ consecutive orbitals. For example,  the root for $1/3$ Laughlin state is  $``1001001\cdots$'' which has at most one electron in each three consecutive orbitals. For electrons with Coulomb interaction at $1/3$ filling, the edge reconstruction happens at $d \sim 1.5 l_B$\cite{Wan02,Wan03} which is signaled by a sudden change of the total angular momentum for ground state as shown in Fig.~\ref{fig1}. Previous numerical studies~\cite{Wan02,Wan03, HuPRB16} show that the spectrum can be perfectly fitted by CLL theory for U(1) bosonic charge mode excitation. The dispersion relation of the edge modes before and after reconstruction for system range from 6 to 10 electrons are shown in Fig.~\ref{fig1}(c) and (d). It shows that the data for different systems sit onto one curve~\cite{HuPRB16, ZhangY14} and the extra edge modes are introduced at $\delta k \neq 0$ which leads to the non-universality of the FQH edge in the tunneling measurements.

\section{Edge reconstruction in real space ES}
\begin{figure}[H]
\centering
\includegraphics[width=6cm]{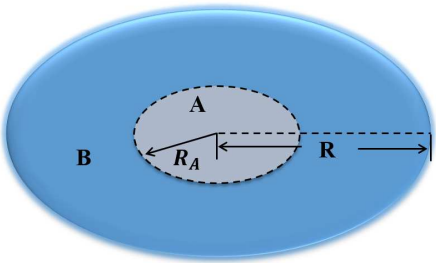}
\caption{\label{fig2} The sketch map of a bipartite finite disk. The real space cut at the radius $R_A$ conserves the rotational symmetry in perpendicular direction.}
\end{figure}
 In this section, we move to the truncated space of the many-body quantum ground state. In order to obtain a bipartite system, a finite disk is divided into two parts as depicting in Fig.~\ref{fig2}. A natural way of splitting the system is the orbital cut since the Hilbert space of the Hamiltonian in Eq.(\ref{Hamiltonian}) is written in the basis of the Landau orbitals. However, the orbital cut appears as a fuzzy cut and not a sharp cut in real space. Several articles~\cite{Regnault12,Dubail12, Simon12} have addressed the question of the real space cut using a sharp real space partition. In this case, the electron operator can be written as
\begin{equation}
 c_m =  \alpha_{m}A_{m} + \beta_{m}B_{m} 
\end{equation}
where $A_m$ and $B_m$ are the operators in A and B subsystem respectively.  $\alpha_{m}^2$($\beta_{m}^2$) is the electron distribution probability of the Landau wave function $|m\rangle$  in part A(B). In particular,
\begin{eqnarray}
\alpha^{2}_m = \int^{R_{A}}_{0}\int^{2\pi}_{0}|\phi_{m}(r,\theta)|^2rdrd\theta =1-\frac{\Gamma(1+m,\frac{R^{2}_{A}}{2})}{\Gamma(1+m)},
\end{eqnarray}
 and $\alpha^{2}_{m}+\beta^{2}_{m}=1$.
The von Neumann entanglement entropy is defined as $S_A = -\text{Tr}[\rho_A \log \rho_A ]$ where $\rho_A = \text{Tr}_B \rho$ is the reduced density matrix of the subsystem after tracing the degrees of freedom in part B. If $\rho_A$ is finite dimensional and has eigenvalues $\lambda_1,\cdots, \lambda_n$, then $S_A = -\sum_i \lambda_i \text{log} \lambda_i$. An alternative way of deriving the entanglement entropy is to perform a Schmidt decomposition of the many-body wave function $|\Psi\rangle = \sum_i e^{-\frac{1}{2}\xi_i} |\psi_A^i\rangle \bigotimes |\psi_B^i\rangle$ giving $\exp(-\frac{1}{2}\xi_i) = \sqrt{\lambda_i}$ as the singular values. Thus the entanglement entropy can be expressed as $S_A = \sum_i \xi_i \exp(-\xi_i)$. It is known that the full structure of the ``entanglement spectrum" (ES) which is the logarithmic Schmidt spectrum of level $\xi_i$ contains much more information about the entanglement between two halves across a cut than the $S_A$.  It plays a key role in analyzing topological order. The structure of the dominant terms in the Schmidt expansion is analogous to the low energy excitations of a many-body Hamiltonian. Especially, for the model wave function such as Eq.(\ref{lauglinwf}), the counting per momentum sector of ES is identical to the energy spectrum of the edge excitation, being due to the bulk-edge correspondence.  Beyond the counting, one could ask whether the entanglement energies of the ES mimics the dispersion of the edge excitation and also has the reconstruction. And if the edge reconstruction happens, what does it tell us for the bulk of the FQH liquid?

\begin{figure}[H]\centering
\includegraphics[width=8.5cm]{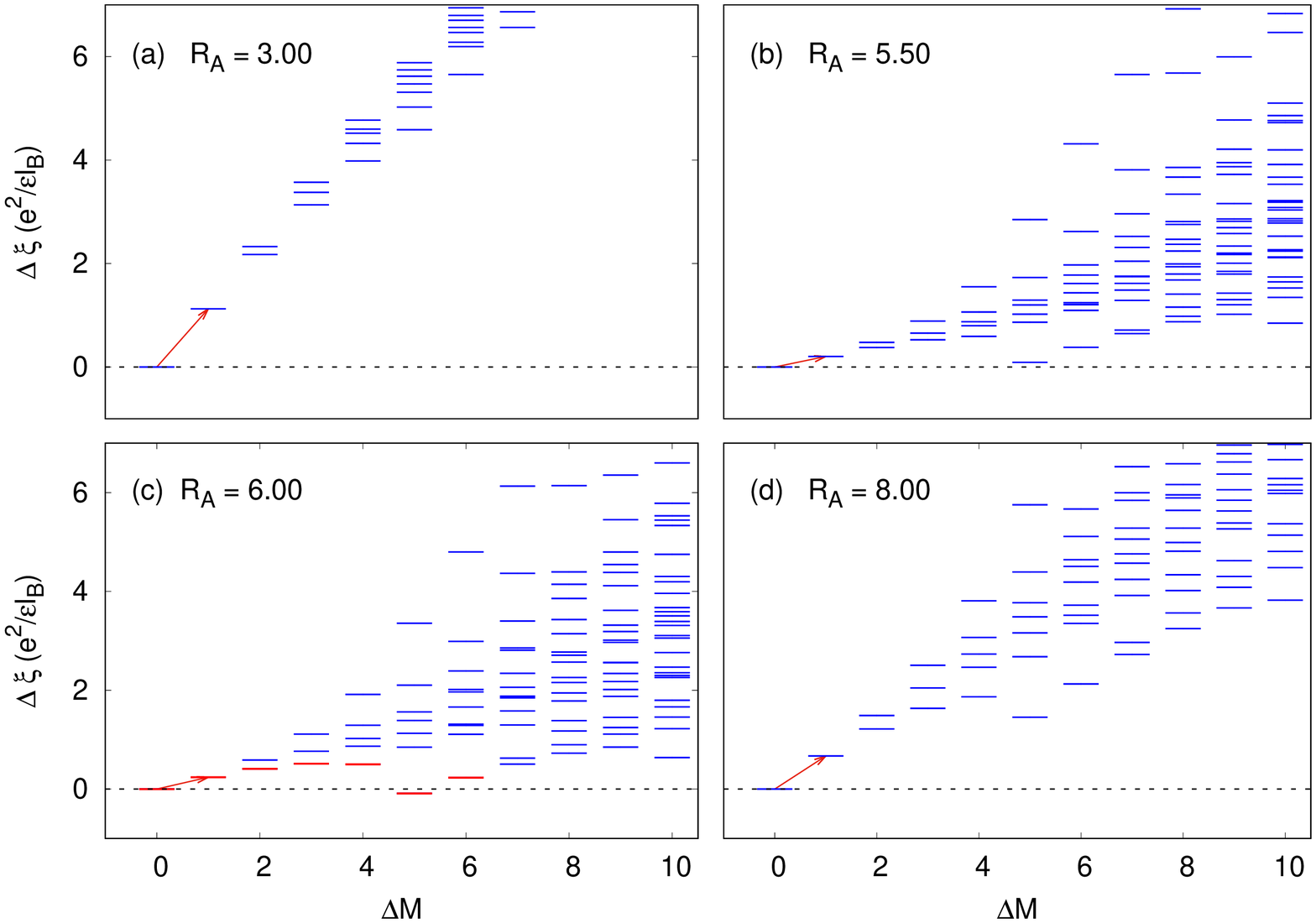}
\caption{\label{fig3} The real space ES for 10 electron Laughlin state with $N_A = 5$ electrons in part A at different cuts.  All the entanglement energies and angular momentum are subtracted by that of the ground state at $M = 30$. The edge reconstruction occurs near $R_A \sim 5.5l_B$ where the energy of the first excited state is almost the same as that of the ground state. The lowest edge states are plotted in red. }
\end{figure}

Hereby, we consider the Laughlin wave function for finite number of electrons on a disc which can be obtained either from diagonalizing a $V_1$ model hamiltonian or from the Jacks. Because the cut conserves the rotational symmetry, in analogy to the energy spectrum in Fig.~\ref{fig1}, the ES for a given number of electrons in subsystem $N_A$ and radius of the circular cut $R_A$ are shown in Fig.~\ref{fig3}. Here we consider a Laughlin state for 10 electrons and the subsystem contains half of the particles. Then the radius of the subsystem $R_A$ is a parameter we tuned. After subtracting the ground state quantum number for subsystem $\Delta M = M_A - \frac{3N_A(N_A-1)}{2}$ in horizontal axis and its entanglement energy $\Delta{\xi_{i}}=\xi_{i}-\xi_{0}$ in vertical axis, as expected the counting number for each momentum subspace is identical to the one in the energy spectrum.  In a finite system, we can then define the edge velocity~\cite{Wan08, Hu09, HuPRB16} through the excitation energy $\Delta \xi(\Delta M = 1)$ of the smallest momentum mode with edge momentum $k = \Delta M / R = 1 / R$, i.e., $v_E = (R/\hbar)\Delta \xi(\Delta M=1)$, where $R_A = \sqrt{2N_A/\nu}$ is the radius of the subsystem. As indicated by arrows in Fig.~\ref{fig3}, the entanglement edge velocity $v_E$ is likely non-monotonic as increasing the radius $R_A$.

\begin{figure} 
\begin{center}
\includegraphics[width=8cm]{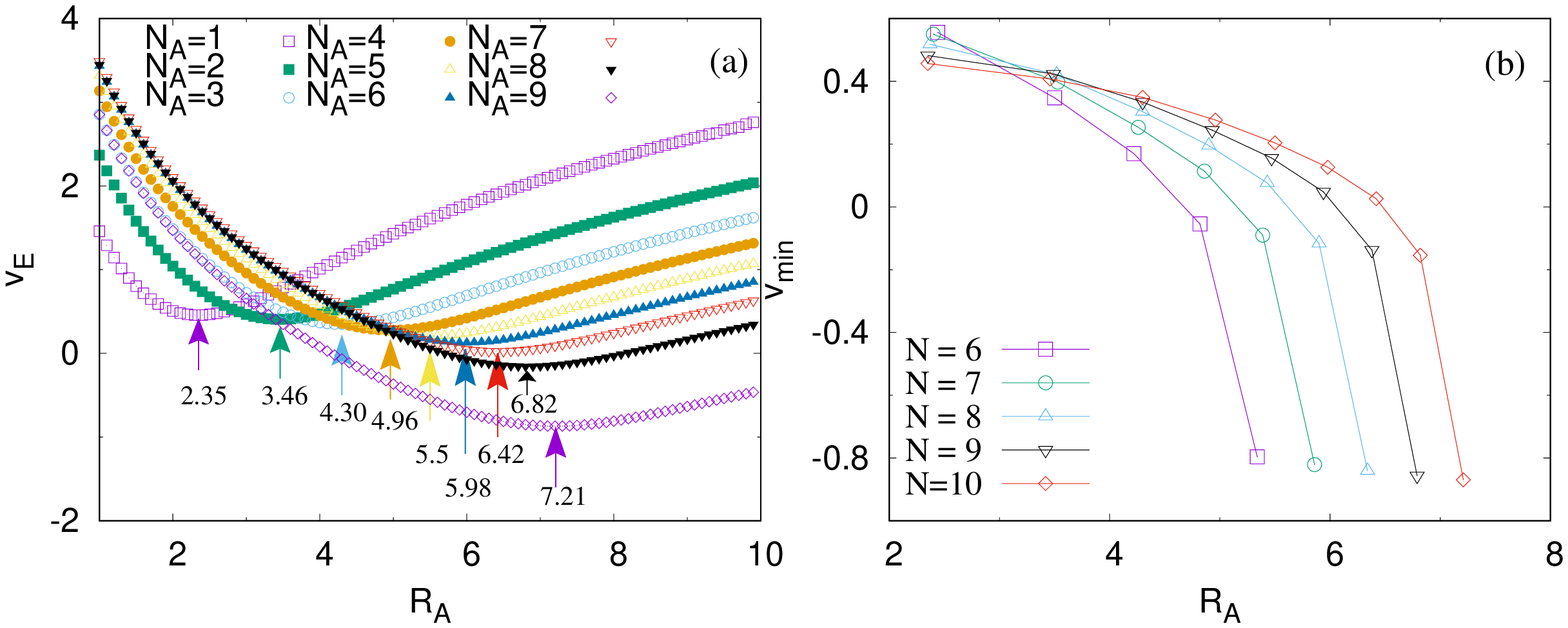}
\caption{\label{fig4} (a) The edge velocities for a given subsystem as a function of $R_A$. The arrows and numbers label the minimum of the $v_E$ for each subsystem. (b) The minimum value of the $v_E$ versus $R_A$ for systems ranging from 6 to 10 electrons. }
\end{center}
\end{figure}

If we plot $v_E$ as a function of $R_A$ for each subsystem with fixed $N_A$, as shown in Fig.~\ref{fig4}(a) for 10 electrons Laughlin state, there is a minimum for each system $N_A$ labelled by arrows.  Interestingly, while $v_E$ taking the value of minimum, the ES seems like having an edge reconstruction.
This is shown in Fig.~\ref{fig3}(b) for subsystem with $N_A = 5$ electrons. When $R_A \sim 5.5$ which is the minima point as shown in Fig.~\ref{fig4}(a), the lowest entanglement energy at $\Delta M = N_A = 5$ is roughly the same as that for the Laughlin state at $\Delta M = 0$.
In Fig.~\ref{fig4}(b), we pick out the minima values of $v_{\text{min}}$ in Fig.~\ref{fig4}(a) and collect all the data from different systems ranging from 6 to 10 electrons. It shows us that the finite size effect becomes small for large system and especially in small subsystem. Therefore, in the thermodynamic limit, 
we believe that the $v_{\text{min}}$ saturates to a fixed value for each subsystem. Another way of defining the edge velocity in the literature is averaging the entanglement energy per angular momentum sector and extrapolating to the thermodynamic limit $N \rightarrow \infty$. It was shown that in this limit the ES dispersion was compatible with a rescaling of the $\nu = 1$ edge mode velocity with a factor $1/\sqrt{3}$~\cite{Regnault}.

 \begin{figure}[H]
\begin{center}
\includegraphics[width=8.5cm]{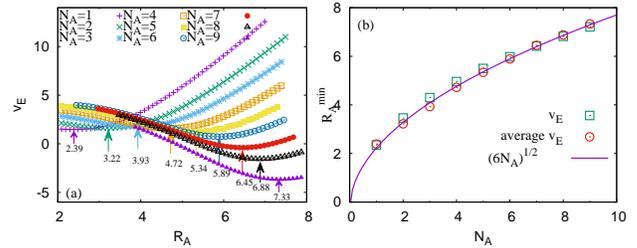}
\caption{\label{fig5}(a) The average edge velocities for a given subsystem as a function of $R_A$. (b) Comparison of the edge velocities between using the lowest two states and the average value in each momentum sector. It shows that the results from both methods satisfy Eq.~(\ref{Rmin}).}
\end{center}
\end{figure}
\begin{figure}[H]
\includegraphics[width=8cm]{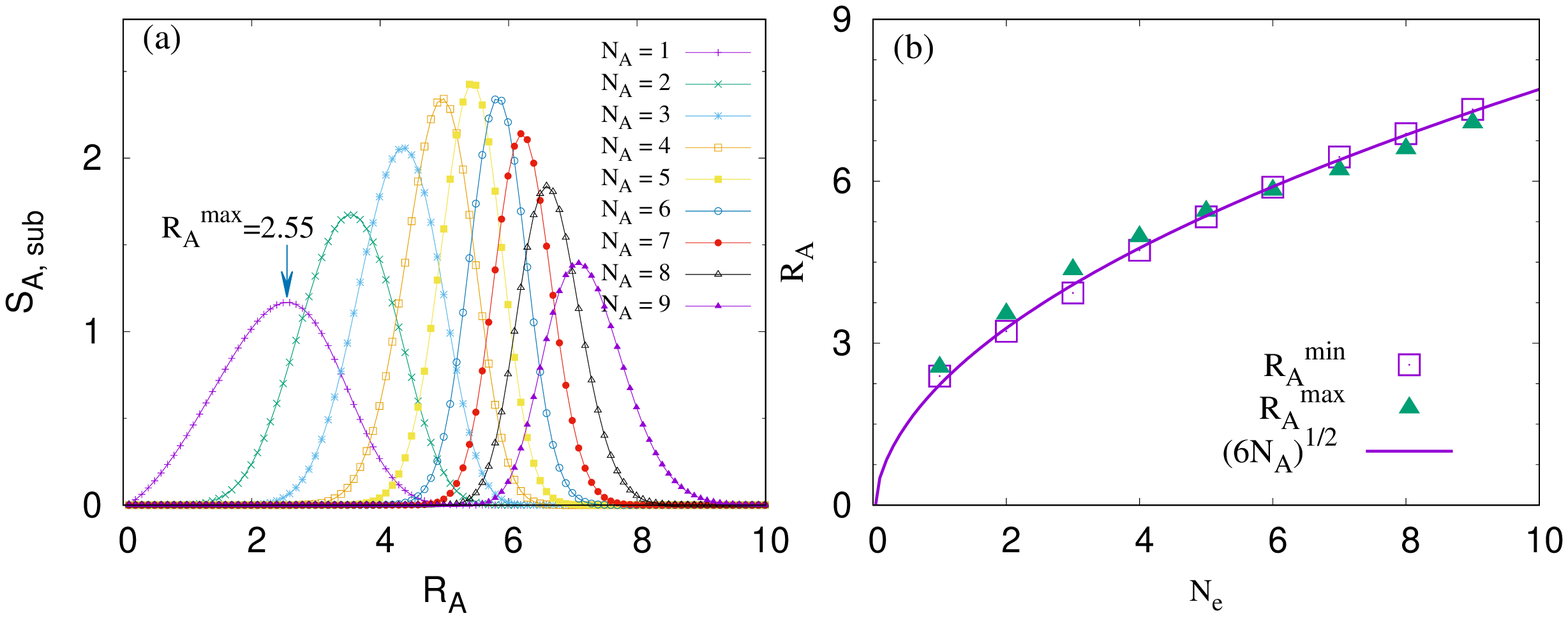}
\caption{\label{fig6} (a) The subsystem entanglement entropy for a 10 electron Laughlin state. The $R_A^{max}$ labels the maximum values of $S_{A, \text{sub}}$. (b) Comparison of the $R_A^{max}$ from the subsystem entanglement entropy and $R_A^{min}$ from minimizing edge velocity. Both are fitted very well by  Eq.~(\ref{Rmin}).}
\end{figure}

In Fig.~\ref{fig5}(a), we plot the averaging velocities in the thermodynamic limit for several subsystems $N_A$. Here the extrapolation is done with data from 6-10 electrons. It is reminiscent of the Fig.~\ref{fig4}(a) that each subsystem still has a minimum which is more or less the same as that from the lowest two states in ES. Similarly, the extrapolation also can be done from Fig.~\ref{fig4}(b) for each  subsystem. In Fig.~\ref{fig5}(b), we plot these extrapolated critical edge velocities from the two methods. As expected, they are highly close to each other and can be fitted very well by 
\begin{eqnarray} \label{Rmin}
R_A^{min} = \sqrt{2N_A/\nu}.
\end{eqnarray}
 The reason is that in the $\nu = 1/3$ Laughlin state, each electron sits in a correlation hole with an area containing three flux quanta. Thus for $N_A$ electrons, there should be $3N_A$ flux quanta to form a correlation hole which corresponding to an circular area with radius $\sqrt{6N_A}$.
 
 According to the above analysis, the real space ES, besides the counting numbers per each momentum subspace, the quantities of the edge spectrum
 still have significant physical meaning. For fixed number of electrons, tuning the radius $R_A$, namely the area of the subsystem can be the engine of the edge reconstruction in ES. The critical area of the subsystem corresponding to the size of the correlation hole for $N_A$ electrons.  To get more intuitions
on the correlation hole, in the following part of this section, we try to obtain the critical radius of the subsystem from another aspect.

 {\it Subsystem entanglement entropy}.  Since the ES and reconstruction we discussed above are restricted in a subsystem with a given $N_A$, it naturally reminds us the the subsystem entanglement entropy, which is the summation of the all the energy levels in ES for a given $N_A$, may have similar critical behaviors near $R_A^{min}$. The subsystem entanglement entropy is defined as the entropy in a system with given $N_A$,
 \begin{eqnarray}
 S_{A, \text{sub}}(N_A, R_A) = \sum_i \xi_{i,N_A} \exp(-\xi_{i, N_A}).
 \end{eqnarray}
The $S_{A, \text{sub}}(N_A, R_A)$ as a function of $R_A$ are depicted in Fig.~\ref{fig6}(a). It shows that the subsystem entanglement entropy  displays a Gaussian-like distribution as a function of $R_A$ which has a maximum point $R_A^{max}$ for each subsystem. Fig.~\ref{fig6}(b) presents the comparison of these maximum and the edge velocity minimum $R_A^{min}$. It tells us that the subsystem entanglement entropy only has nonzero contribution near the edge of the correlation hole and is maximized exactly at the edge, or the correlation hole only has correlation (or entanglement) at the edge. Here we want to remind that the subsystem entanglement entropy is a partial entanglement entropy which can not be used to apply the ``area law" or extrapolate the topological entanglement entropy. The entanglement entropy should be the summation of the $S_{A, \text{sub}}(N_A, R_A)$ for all the subsystems $N_A \in [0, N_e]$.  

\section{Coulomb interaction and other FQH states}

\begin{figure}
\begin{center}
\includegraphics[width=8cm]{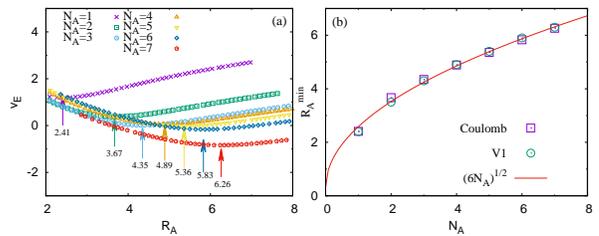}
\caption{\label{fig7} The entanglement edge velocities as a function of $R_A$ for Coulomb interaction (a) and its minimum values compared with that of the model wave function labelled by $V_1$. Again, both are fitted very well by Eq.~(\ref{Rmin}).}
\end{center}
\end{figure}

 {\it Coulomb interaction}.  The analysis in previous section is for the model wave function. Here we double check the validity of the conclusions in the case of realistic Coulomb interaction.  Here we fix the background confinement at $d = 1.0l_B$ and calculate the real space ES for each subsystem because the background potential has
 insignificant effects on the wave function of a gapped topological ground state. In reminiscent of that for the model wave function, the entanglement velocity for Coulomb interaction still has a minimum while increasing the radius of circular cut $R_A$ as depicted in Fig.~\ref{fig7}(a). The arrows still label the minima points of the velocities. In Fig.~\ref{fig7}(b), we compare these results with that from the model wave function via diagonalizing $V_1$ Hamiltonian. It is shown that the Coulomb interaction is consistent exactly with the model Hamiltonian and again, obeys the relation $R_A^{min} = \sqrt{2N_A/\nu}$.

\begin{figure}
\includegraphics[width=8cm]{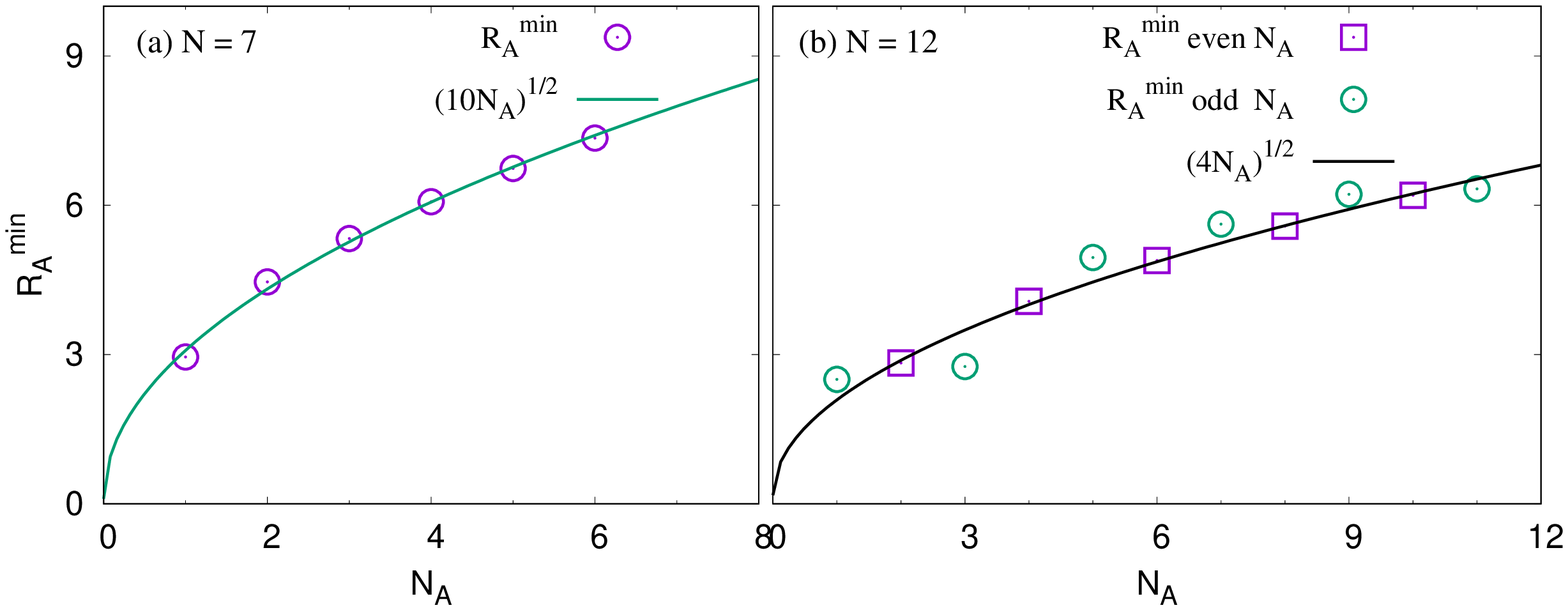}
\caption{\label{fig8} By calculating the entanglement edge velocities and locating its minimum, the scaling of the $R_A^{min}$ for $1/5$ Laughlin state (a) and Moore-Read state (b) still satisfy the Eq.~(\ref{Rmin}). The Moore-Read state has a strong even-odd effect in the edge velocity which demonstrates the pairing properties of this non-Abelian states.}
\end{figure}
 {\it Other FQH states}. In the following, we consider the other two typical FQH states. The  Laughlin state at $\nu = 1 /5$ and the non-Abelian Moore-Read state which is the candidate wave function for $\nu = 5/2$ FQH state.~\cite{MR}
\begin{eqnarray}\label{lauglinMR}
\Psi_L^5(\{z_i\}) &=&\prod_{i<j}(z_i-z_j)^5 e^{-\frac{1}{4}\sum_i |z_i|^2} \\
\Psi_{MR}(\{z_i\})&=& \text{Pf}(\frac{1}{z_i-z_j}) \prod_{i<j}(z_i-z_j)^2 e^{-\frac{1}{4}\sum_i |z_i|^2}
\end{eqnarray}
The $1/5$ Laughlin wave function is the model wave function for the two-body $\{V_1, V_3\}$ model Hamiltonian and the Moore-Read Pfaffian wave function is the model wave function for the three-body $V_3$ model Hamiltonian. Of course, both of them can also be obtained from the Jacks.  In Fig.~\ref{fig8}, we present the minima edge velocity for different subsystems in two FQH states.  In reminiscent of that in the $1/3$ Laughlin state, the results for $1/5$ Laughlin state scale very well with the relation
$R_A^{min} = \sqrt{10N_A}$ which demonstrates that the correlation hole of the $1/5$ Laughlin state contains five flux quanta.  The results for Moore-Read state are shown in Fig.~\ref{fig8}(b).  Interestingly, from the macroscopic level, the $R_A^{min} = \sqrt{4N_A}$ is satisfied which means two flux quanta for each correlation hole in the Pfaffian state. Especially, the data for the even number of $N_A$ exactly match this relation. However, the data for subsystem with odd number of electrons has obvious deviations. It definitely shows us the pairing mechanism of the Pfaffian state, namely the correlation hole of the Moore-Read state should be four flux quanta occupied by two electrons. This is the fundamental unit of the Pfaffian state which labelled by ``1100" in the root of Jack polynomial.

\section{ Conclusions and discussions}
In conclusion, we systematically studied the properties of the real space entanglement spectrum, especially the entanglement velocity in case we treat the ES as the edge excitation near the bipartite cut. On a finite disc,  in reminiscent of the edge reconstruction of the energy spectrum while smoothing the strength of the background confinement, the ES for a given subsystem with fixed number of electrons $N_A$ can also have edge reconstruction while increasing the area of the subsystem.  The ES edge reconstruction corresponds to the minimum of the edge velocity. In the case of a circular cut which conserves the rotational symmetry, the corresponding radius of the subsystem satisfies the relation $R_A^{min} = \sqrt{2N_A/\nu}$ which demonstrates the size of the fundamental correlation hole in FQH states. Besides the $1/3$ Laughlin wave function, this conclusion is also supported by the realistic Coulomb interaction and other FQH model wave functions, such as $1/5$ Laughlin state and the Moore-Read Pfaffian state. We thus conclude that the ES can tells us not only the counting number of the CFT edge states, but also much more the physical properties of the FQH liquids, such as the edge reconstruction and size of the fundamental correlation hole.  Especially, the strong even-odd effects of the $R_A^{min}$ for the Moore-Read wave function demonstrates the pairing property of the state.

Here we should note that our study of the real space ES is based on the circular cut in the bulk. This is because we are studying the isotropic FQH states which conserves the rotational symmetry in disk geometry.  For the generalized isotropic FQH states which do not conserve the rotation symmetry~\cite{Qiu12}, we believe the real space cut should follows the geometric shape of the Landau orbitals. Moreover, for a general anisotropic FQH state which breaks both the rotational and translational symmetries~\cite{Haldane11}, in spite of that we can not plot the ES in the momentum sector and explore the edge reconstruction without rotational symmetry, it is worth to explore relation of the subsystem entanglement entropy and the intrinsic metric in the correlation hole.~\cite{Ye}

\acknowledgements
This work is supported by National Natural Science Foundation of China Grants No.11674041, 11974064, Chongqing Research Program of Basic Research and Frontier Technology Grant No. cstc2017jcyjAX0084. Q. Li thanks for support of the National Natural Science Foundation of China Grant No.11474144.  N. Jiang is supported by the National Natural Science Foundation of China Grant No. 11674282 and the Strategic Priority Research Program of Chinese Academy of Sciences, Grant No. XDB28000000.

\end{document}